\documentclass[a4paper,11pt]{article}

\usepackage{jheppub}

\usepackage{amsfonts,graphics,epsfig,subfigure}

\title{Phase transitions, geometrothermodynamics and critical exponents of black holes with conformal anomaly}

\author[a,b]{Jie-Xiong Mo}
\author[a,1]{Wen-Biao Liu \note{Corresponding author}}

\affiliation[a]{Department of Physics, Institute of Theoretical
Physics, Beijing Normal University,\\Beijing, 100875, China}
\affiliation[b]{Institute of Theoretical Physics, Zhanjiang Normal
University,\\Zhanjiang, Guangdong, 524048, China}

\emailAdd{mojiexiong@gmail.com}
\emailAdd{wbliu@bnu.edu.cn}

\abstract{Conformal anomaly is an
important concept which has various applications in
quantum field theory in curved space-time, string theory, black hole
physics and cosmology. Probing its influences in phase
transitions of black holes is of great physical importance. In this paper, we achieve this goal by investigating the phase transitions of black holes with conformal
anomaly in canonical ensemble from different
perspectives. Some interesting and novel phase transition phenomena has been discovered. Firstly, we discuss the behavior of the specific heat and the inverse
of the isothermal compressibility. It is shown that there are striking differences in
Hawking temperature and phase structure between black holes with conformal anomaly and those without it. In the case with conformal
anomaly, there exists local minimum temperature corresponding to
the phase transition point; Phase transitions take place not only from an unstable large black
hole to a locally stable medium black hole but also from an unstable
medium black hole to a locally stable small black hole. Secondly, we probe in details the dependence of
phase transitions on the choice of parameters. The results show that
black holes with conformal anomaly have much richer phase structure
than those without it. There would be two, only one
or no phase transition points depending on the parameters we have
chosen. The corresponding parameter region are derived both numerically and graphically. Thirdly, geometrothermodynamics are built up to examine the phase structure we have discovered. It is shown that Legendre invariant thermodynamic scalar
  curvature diverges exactly where the specific heat diverges. Furthermore, critical behaviors are investigated by calculating the relevant critical exponents.
And we proved that these critical exponents satisfy the
thermodynamic scaling laws, leading to the conclusion that critical
exponents and the scaling laws do not change even when we consider
conformal anomaly.}

\begin{document}
\maketitle
\flushbottom

\section{Introduction}

   Black hole thermodynamics has long been one of exciting and challenging research fields ever since the pioneer research made by Bekenstein and
     Hawking~\cite{Bekenstein}-\cite{Hawking1}. A variety of thermodynamic quantities of black holes has been studied. In 1983, Hawking and
     Page~\cite{Hawking2} discovered that pure thermal radiation in AdS space becomes
unstable above certain temperature and collapses to form black
holes. This is the well-known Hawking-Page phase transition which
describes the phase transition between the Schwarzschild AdS black
hole and the thermal AdS space. This phenomenon can be interpreted
in the AdS/CFT correspondence~\cite{Maldacena9999} as the
confinement/deconfinement phase transition of gauge
field~\cite{Witten9999}. Since then, phase transitions of black
holes have been investigated from different perspectives. For recent
papers, see~\cite{Sahay}-\cite{Wenbiao1}.

 One of the elegant approach is the thermodynamic geometry method, which was first introduced by Weinhold \cite{Weinhold} and Ruppeiner \cite{Ruppeiner}. Weinhold proposed metric structure in the energy representation as $g_{i,j}^{W}=\partial_{i}\partial_{j}M(U,N^a)$ while Ruppeiner defined metric structure
    as $g_{i,j}^{R}=-\partial_{i}\partial_{j}S(U,N^a)$. These metric structures are respectively the Hessian matrix of the internal energy $U$  and the entropy $S$
    with respect to the extensive thermodynamic variables $N^a$. And Weinhold's metrics were found to be conformally connected to Ruppeiner's metrics through the map
    $dS^2_R=\frac{dS^2_W}{T}$~\cite{Janyszek}. Ruppeiner's metric has been applied to investigate various thermodynamics systems for its profound physical meaning. For more details, see the nice review~\cite{Ruppeiner2}. Recently Quevedo et al. \cite{Quevedo2} presented a new formalism called geometrothermodynamics, which allows us to derive Legendre invariant metrics in the space of equilibrium states. Geometrothermodynamics  presents a unified geometry where the metric structure describes various types of black hole thermodynamics \cite{Quevedo3}-\cite{Wenbiao2}.

Apart from the thermodynamic geometry, critical behavior also plays
a crucial role in the study of black hole phase transitions. The
critical points of phase transitions are characterized by the
discontinuity of thermodynamic quantities. So it is important to
investigate the behavior in the neighborhood of the critical point,
especially the divergences of various thermodynamic quantities. In
classical thermodynamics, this goal is achieved by taking into
account a set of critical exponents, from which we can gain
qualitative insights into the critical behavior. These critical
exponents are found to be universal to a large extent (only
depending on the dimensionality, symmetry etc) and satisfy scaling
laws, which can be attributed to scaling hypothesis. Critical
behavior of black holes accompanied with their critical exponents
have been investigated not only in asymptotically flat space time
\cite{Davies}-\cite{Arcioni1} but also in the de Sitter and anti de
Sitter space \cite{Muniain}-\cite{Liu99}.

    In this paper, we would like to focus our attention on the critical behavior and
geometrothermodynamics of static and spherically symmetric black
holes with conformal anomaly. As we know, conformal anomaly, an
important concept with a long history, has various applications in
quantum field theory in curved spaces, string theory, black hole
physics and cosmology. So it is worth probing its influences in phase
transitions of black holes. Recently, Cai et al.~\cite{Cai9999}
found a class of static and spherically symmetric black holes with
conformal anomaly, whose thermodynamic quantities were also
investigated in the same paper. It was found that there exists a
logarithmic correction to the well-known Bekenstein-Hawking area
entropy. This discovery is quite important in the sense that with
this term one is able to compare black hole entropy up to the
sub-leading order, in the gravity side and in the microscopic
statistical interpretation side~\cite{Cai9999}. Based on the metrics
in that paper, phase transitions of a spherically symmetric
Schwarzschild black hole have been investigated by taking into
account the back reaction through the conformal anomaly of matter
fields recently~\cite{Son9999}. It has been shown that there exists
an additional phase transition to the conventional Hawking-Page
phase transition. The entropy of these black holes has also been investigated by using quantum tunneling approach~\cite{Liran}. Moreover, Ehrenfest equation has been applied
to investigate this class of black holes~\cite{Chenghongbo} and it
has been found that the phase transition is a second order one.
Despite of these achievements, there are still many issues left to
be explored. Ref.~\cite{Son9999} mainly focus on the uncharged case
. So it is natural to ask what would happen to the charged black
holes. Ref.~\cite{Chenghongbo} concentrated their efforts on the
Ehrenfest equation in the grand canonical ensemble. So it is
worthwhile to study the phase transition in canonical ensemble. The
dependence of the phase structure on the parameter deserves to be
further investigated. One may also wonder whether the thermodynamic
geometry and scaling laws still works to reveal the phase structure
and critical behavior when conformal anomaly is taken into
consideration. Motivated by these, we would like to investigate the
phase transition, geometrothermodynamics and critical exponents in
canonical ensemble.

    The organization of our paper is as follows. In Sec.~\ref{sec:2}, the thermodynamics of
    black holes with conformal anomaly will be briefly reviewed. In Sec.~\ref
{sec:3}, phase transitions will be investigated in details in
canonical ensemble and some interesting and novel phase transition phenomena will be disclosed. In Sec.~\ref {sec:4}, geometrothermodynamics
will be established to examine the phase structure we find in Sec.~\ref {sec:3}. In
Sec.~\ref {sec:5}, critical exponents will be calculated and the
scaling laws will be examined. In the end, conclusions will be drawn
in Sec.~\ref {sec:6}.

\section{A brief review of thermodynamics}
\label{sec:2} The static and spherically symmetric black hole
solution with conformal anomaly has been proposed as~\cite{Cai9999}
\begin{equation}
ds^2=f(r)d t^2-\frac{d r^2}{f(r)}-r^2(d \theta^2+sin^2 \theta d
\varphi^2),\label{1}
\end{equation}
where
\begin{equation}
f(r)=1-\frac{r^2}{4\tilde{\alpha}}(1-\sqrt{1-\frac{16\tilde{\alpha}
M}{r^3}+\frac{8\tilde{\alpha} Q^2}{r^4}}\,).\label{2}
\end{equation}
The Newton constant $G$ has been set to one. Both $M$ and $Q$ are
integration constants. And the coefficient $\tilde{\alpha}$ is
positive. The physical meanings of $M$ and $Q$ were discussed in
Ref.~\cite{Cai9999}. $M$ is nothing but the mass of the solution
while $Q$ should be interpreted as $U(1)$ charge of some conformal
field theory.

When $M=Q=0$, the metric above reduces to
\begin{equation}
ds^2=d t^2-d r^2-r^2(d \theta^2+sin^2 \theta d \varphi^2),\label{3}
\end{equation}
implying that the vacuum limit is the Minkowski space-time.

 In the large $r$ limit, Eq.({\ref{2}}) becomes
 \begin{equation}
f(r)\approx1-\frac{2M}{r}+\frac{Q^2}{r^2}+O(r^{-2}),\label{4}
\end{equation}
which behaves like the Reissner-Norstr\"{o}m solution.

When $\tilde{\alpha}\rightarrow0$, Eq.({\ref{2}}) reduces into
 \begin{equation}
f(r)=1-\frac{2M}{r}+\frac{Q^2}{r^2},\label{5}
\end{equation}
Eqs.({\ref{1}}) and ({\ref{5}}) consist of the metric of
Reissner-Norstr\"{o}m black hole.

 Solving the
equation $f(r)=0$, we can get the radius of black hole horizon
$r_+$, with which the mass of the black hole can be expressed as
\begin{equation}
M=\frac{r_+}{2}+\frac{Q^2}{2r_+}-\frac{\tilde{\alpha}}{r_+}.\label{6}
\end{equation}
The Hawking temperature can be derived as
\begin{equation}
T=\frac{f'(r_+)}{4\pi}=\frac{r_+^2+2\tilde{\alpha}-Q^2}{4\pi
r_+(r_+^2-4\tilde{\alpha})}.\label{7}
\end{equation}
The potential difference between the horizon and the infinity can be
written as
\begin{equation}
\Phi=\frac{Q}{r_+}.\label{8}
\end{equation}
 The entropy was reviewed in Ref.~\cite{Chenghongbo}
as
\begin{equation}
S=\pi r_+^2-4\pi \tilde{\alpha}ln{r_+^2}.\label{9}
\end{equation}

\section{Novel phase transition phenomena}
\label{sec:3} In this section, we would like to investigate the
phase transition of black holes with conformal anomaly in canonical
ensemble where the charge of the black hole is fixed.

\begin{figure*}
\centerline{\subfigure[]{\label{1a}
\includegraphics[width=8cm,height=6cm]{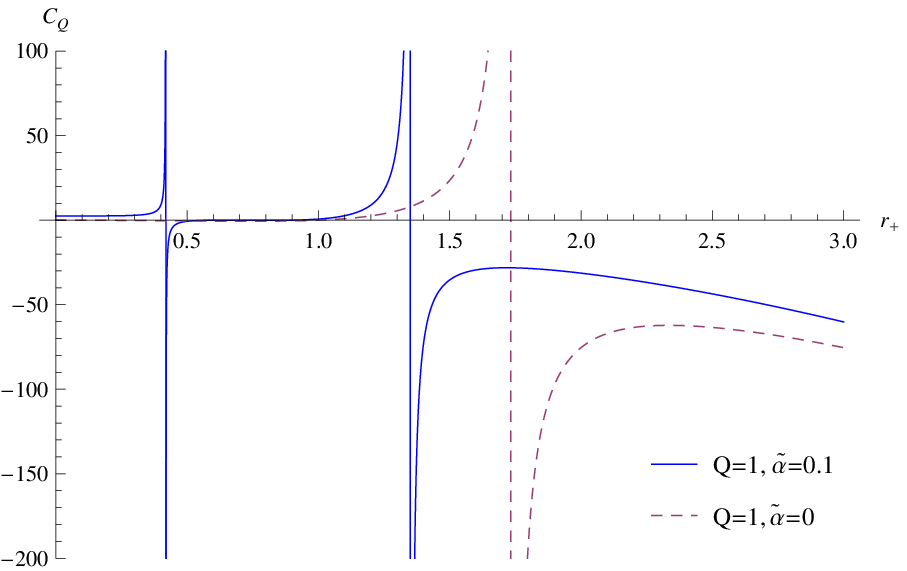}}
\subfigure[]{\label{1b}
\includegraphics[width=8cm,height=6cm]{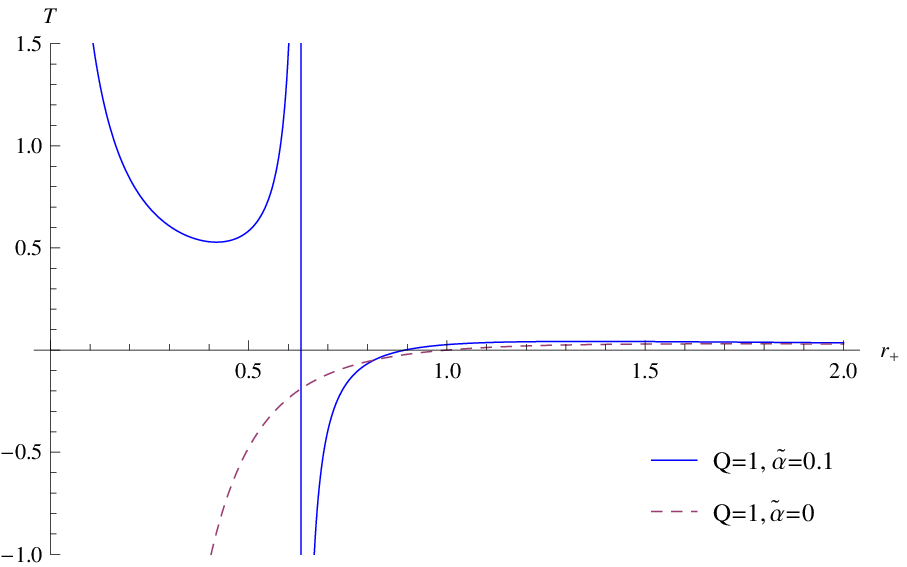}}}
 \caption{(a)$C_Q$ vs. $r_+$ for
 $Q=1,\tilde{\alpha}=0.1$ (b)$T$ vs. $r_+$ for
 $Q=1,\tilde{\alpha}=0.1$} \label{fg1}
\end{figure*}

 The corresponding specific heat can be calculated as
\begin{equation}
C_Q=T(\frac{\partial S}{\partial
T})_Q=\frac{2\pi(r_+^2-4\tilde{\alpha})^2(Q^2-2\tilde{\alpha}-r_+^2)}{r_+^4-8\tilde{\alpha}^2+10r_+^2\tilde{\alpha}
+Q^2(4\tilde{\alpha}-3r_+^2)}.\label{10}
\end{equation}
Apparently, $C_Q$ may diverge when
\begin{equation}
r_+^4-8\tilde{\alpha}^2+10r_+^2\tilde{\alpha}
+Q^2(4\tilde{\alpha}-3r_+^2)=0,\label{11}
\end{equation}
which suggests a possible phase transition. However, the phase
transition point characterized by Eq.(\ref{11}) is not apparent. To
gain an intuitive understanding, we plot Fig.\ref{1a} using
Eq.(\ref{10}). To check whether the phase transition point locates
in the physical region, we also plot the Hawking temperature using
Eq.(\ref{7}) in Fig.\ref{1b}. It is shown that the phase transition
point locates in the positive temperature region. From Fig.\ref{1a}
and Fig.\ref{1b}, we find that there have been striking differences
between the case $\tilde{\alpha}\neq0$ and the case
$\tilde{\alpha}=0$. In the case $Q=1,\tilde{\alpha}=0.1$, there are
two phase transition point while there is only one in the case
$\tilde{\alpha}=0$. The temperature in the case $\tilde{\alpha}=0$
increases monotonically while there exists local minimum temperature
in the case $Q=1,\tilde{\alpha}=0.1$. Fig.\ref{1a} can be divided
into four phases. The first one is thermodynamically stable ($C_Q>
0$)with small radius. The second one is unstable ($C_Q< 0$)with
meidium radius. The third one is thermodynamically stable ($C_Q>
0$)with medium radius while the fourth one is thermodynamically
unstable ($C_Q<0$)with large radius. So the phase transition take
place not only from an unstable large black hole to a locally stable
medium black hole but also from an unstable medium black hole to a
locally stable small black hole.
From Fig.\ref{1b},we notice that the Hawking temperature has a local
minimum value. And the corresponding location can be derived through
\begin{equation}
\frac{\partial T}{\partial
r_+}=-\frac{r_+^4-8\tilde{\alpha}^2+10r_+^2\tilde{\alpha}
+Q^2(4\tilde{\alpha}-3r_+^2)}{4\pi (r_+^3-4r_+\tilde{\alpha})^2}=0.
\label{12}
\end{equation}
It is quite interesting to note that the numerator of Eq.(\ref{12})
is the same as Eq.(\ref{11}), which implies that the location which
corresponds to the minimum Hawking temperature also witnesses the
existence of phase transition.

 To probe the dependence of phase transition location on the
   choice of parameter, we solve Eq.(\ref{11}) and obtain the
   location of phase transition point as
\begin{equation}
r_c=\sqrt{\frac{3Q^2-10\tilde{\alpha}\pm\sqrt{132\tilde{\alpha}^2-76\tilde{\alpha}
Q^2+9Q^4}}{2}}.\label{13}
\end{equation}

\begin{figure*}
\centerline{\subfigure[]{\label{2a}
\includegraphics[width=8cm,height=6cm]{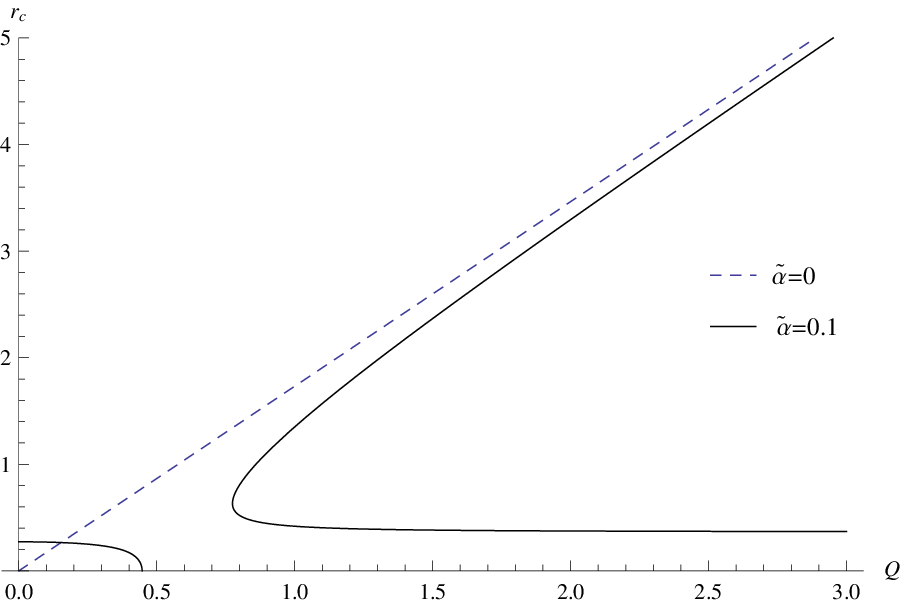}}
\subfigure[]{\label{2b}
\includegraphics[width=8cm,height=6cm]{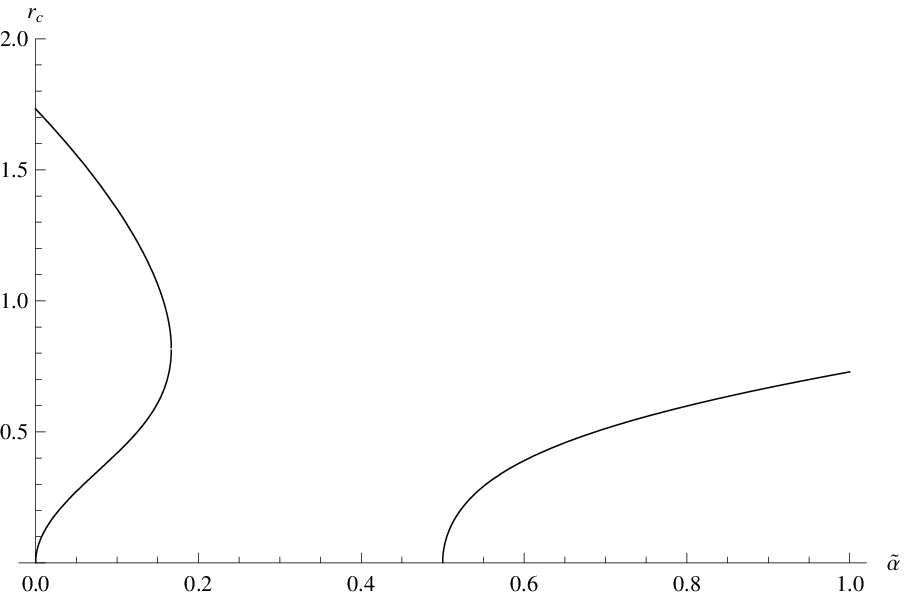}}}
 \caption{(a)$r_c$ vs. $Q$ for
 $\tilde{\alpha}=0.1$ and  $\tilde{\alpha}=0$ (b)$r_c$ vs. $\tilde{\alpha}$ for
 $Q=1$} \label{fg2}
\end{figure*}

\begin{figure*}
\centerline{\subfigure[]{\label{3a}
\includegraphics[width=8cm,height=6cm]{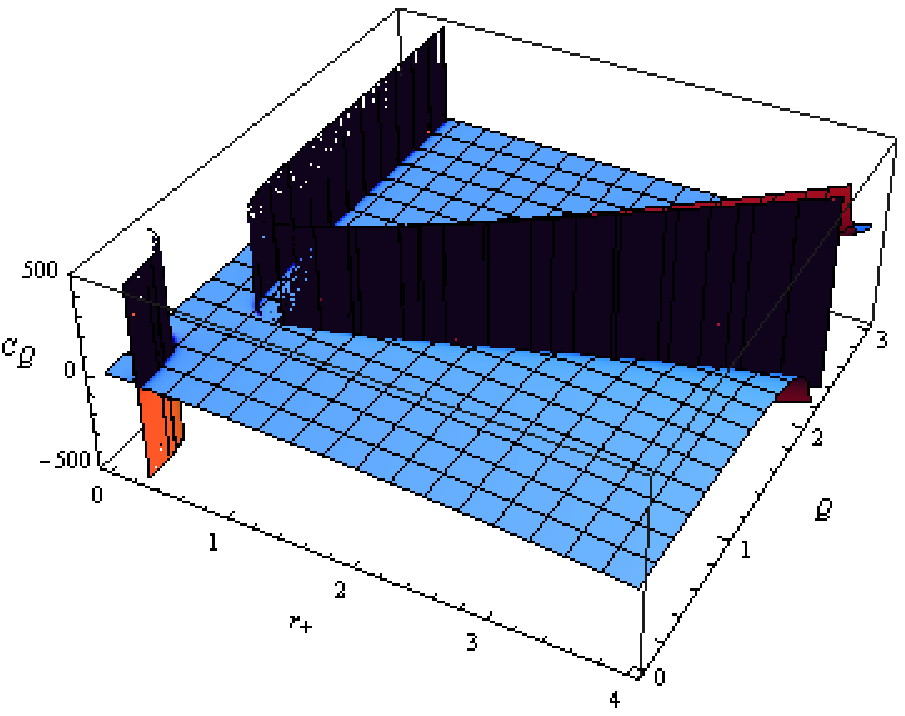}}
\subfigure[]{\label{3b}
\includegraphics[width=8cm,height=6cm]{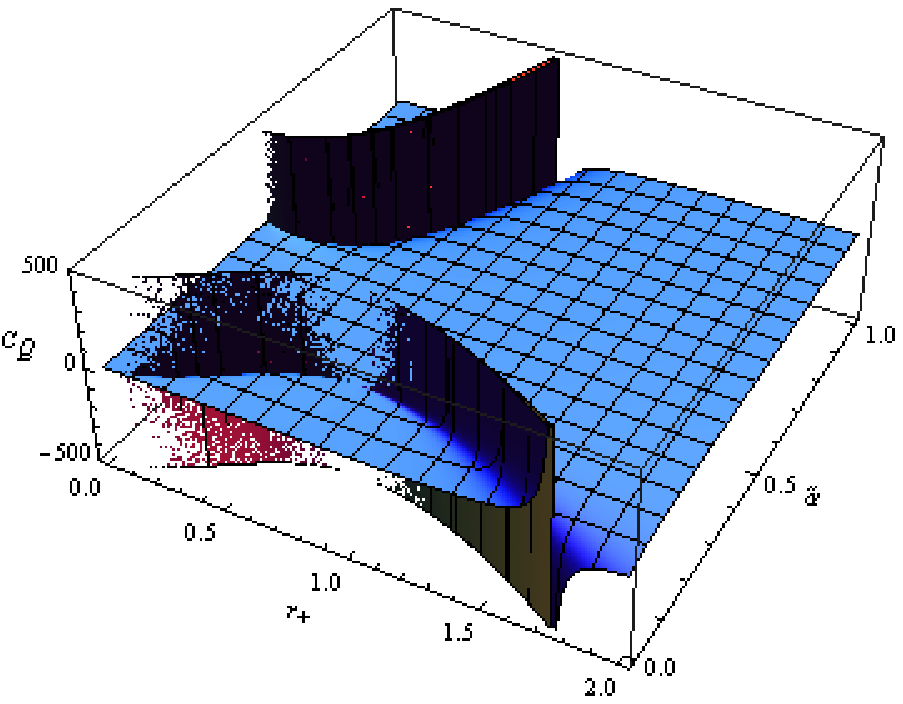}}}
 \caption{(a)$C_Q$ vs. $Q$ and $r_+$ for
 $\tilde{\alpha}=0.1$ (b)$C_Q$ vs. $\tilde{\alpha}$ and $r_+$ for $Q=1$}   \label{fg3}
\end{figure*}

With Eq.(\ref{13}) at hand, we plot Fig.\ref{2a} and Fig.\ref{2b}
which show the influence of parameters $Q$ and $\tilde{\alpha}$
respectively.
It can be observed from Fig.\ref{2a} and Fig.\ref{2b} that black
holes with conformal anomaly have much richer phase structure than
that without conformal anomaly. When $\tilde{\alpha}=0$, the
location of the phase transition $r_c$ is proportional to the charge
$Q$. However, the cases of black holes with conformal anomaly are
quite complicated. For $\tilde{\alpha}=0.1$, the curve can be
divided into three regions. Through numerical calculation, we find
that black holes have only one phase transition point when
$Q\subset(0,0.4472)$. When $0.4472<Q<0.7746$, there would be no
phase transition at all. When $Q>0.7746$, there exist two phase
transition points, just as what we show in Fig.\ref{1a}. And the
distance between these two phase transition point becomes larger
with the increasing of $Q$. Fig.\ref{2b} shows the case that the
charge $Q$ has been fixed at one. We notice that there would be two
phase transition points when $0<\tilde{\alpha}<\frac{1}{6}$, which is
consistent with Fig.\ref{1a}. And the distance between these two
phase transition point becomes narrower with the increasing of
$\tilde{\alpha}$. When
$\tilde{\alpha}\subset(\frac{1}{6},\frac{1}{2})$, there would be no
phase transition. When $\tilde{\alpha}>\frac{1}{2}$, there would be
only one phase transition point. To gain a three-dimensional
understanding, we also include a three dimensional figure of $C_Q$
in Fig.\ref{3a} and Fig.\ref{3b}.

Apart from the specific heat, we would also like to investigate the
behavior of the inverse of the isothermal compressibility, which is
defined as

\begin{equation}
\kappa _T^{-1}=Q(\frac{\partial\Phi}{\partial Q})_T.\label{14}
\end{equation}
Utilizing the thermodynamic identity relation
\begin{equation}
(\frac{\partial\Phi}{\partial T})_Q(\frac{\partial T}{\partial
Q})_\Phi(\frac{\partial Q}{\partial \Phi})_T=-1,\label{15}
\end{equation}
we obtain
\begin{equation}
(\frac{\partial\Phi}{\partial Q})_T=-(\frac{\partial\Phi}{\partial
T})_Q(\frac{\partial T}{\partial Q})_\Phi,\label{16}
\end{equation}
where the second term on the right hand side can be calculated
through
\begin{equation}
(\frac{\partial T}{\partial Q})_\Phi=(\frac{\partial T}{\partial
r_+})_Q(\frac{\partial r_+}{\partial Q})_\Phi+(\frac{\partial
T}{\partial Q})_{r_+}.\label{17}
\end{equation}
Utilizing Eqs.(\ref{7}), (\ref{8}), (\ref{14}), (\ref{16}),
(\ref{17}), we obtain the explicit form of $\kappa _T^{-1}$ as
\begin{equation}
\kappa
_T^{-1}=\frac{Qr_+^4-Q^3r_+^2-4Q^3\tilde{\alpha}+10Qr_+^2\tilde{\alpha}-8Q\tilde{\alpha}^2}{r_+[r_+^4-8\tilde{\alpha}^2+10r_+^2\tilde{\alpha}
+Q^2(4\tilde{\alpha}-3r_+^2)]}.\label{18}
\end{equation}

\begin{figure}
\includegraphics[width=8cm,height=6cm]{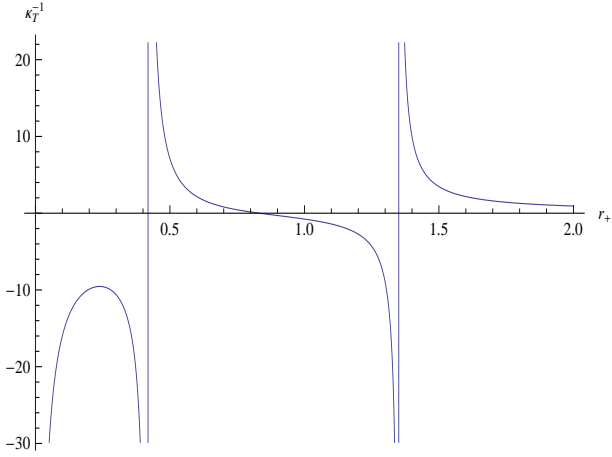}
 \caption{The inverse of the isothermal compressibility $\kappa _T^{-1}$ vs. $r_+$ for $Q=1,\tilde{\alpha}=0.1$}
\label{fg4}
\end{figure}

We show the behavior of $\kappa_T^{-1}$ in Fig.\ref{fg4}. Comparing
Fig.\ref{fg4} with Fig.\ref{1a}, we find that the inverse of the
isothermal compressibility $\kappa_T^{-1}$ also diverges at the
critical point.

\section{Geometrothermodynamics}
\label{sec:4}
    According to geometrothermodynamics~\cite{Quevedo2}, the $(2n+1)$-dimensional thermodynamic phase space $\mathcal {T}$ can be coordinated by the set of independent quantities \{$\phi,E^a,I^a$\},
     where $\phi$ corresponds to the thermodynamic potential, and $E^a,I^a$ are the extensive and intensive thermodynamic variables respectively. The fundamental Gibbs 1- form defined on $\mathcal {T}$ can then be written as $\Theta=d\phi-\delta_{ab}I^adE^b$, where $\delta_{ab}=diag(1,\cdots,1)$.
      Considering a non-degenerate Riemannian metric $G$, a contact Riemannian manifold can be defined from the set $(\mathcal {T},\Theta,G)$ if the
      condition $\Theta\wedge(\Theta)^n\neq0$ is satisfied. Utilizing a smooth map $\varphi:\varepsilon\rightarrow\mathcal{T}$, i.e. $\varphi:(E^a)\mapsto(\phi,E^a,I^a)$,
      a submanifold $\varepsilon$ called the space of thermodynamic equilibrium states can be induced. Furthermore, a thermodynamic metric $g$ can be induced in
      the equilibrium manifold $\varepsilon$ by the smooth map $\varphi$.

          As proposed by Quevedo, the non-degenerate metric $G$ and the thermodynamic metric $g$ can be written as
          follows~\cite{Quevedo7}
\begin{equation}
G=(d\phi-\delta_{ab}I^adE^b)^2+(\delta_{ab}E^aI^b)(\eta_{cd}dE^cdI^d),\label{19}
\end{equation}%
 \begin{equation}
g=\varphi^*(G)=(E^c\frac{\partial \phi}{\partial
E^c})(\eta_{ab}\delta^{bc}\frac{\partial^2\phi}{\partial E^c
\partial E^d}dE^adE^d),\label{20}
\end{equation}%
where $\eta_{ab}=diag(-1,\cdots,1)$.

   To construct geometrothermodynamics of black holes with conformal anomaly in canonical ensemble, we
choose $M$ to be the thermodynamic potential and $S,Q$ to be the
extensive variables. Then the corresponding thermodynamic phase
space is a 5-dimensional one coordinated by the set of independent
coordinates\{$M, S, Q, T, \Phi$\}. The fundamental Gibbs 1- form
defined on $\mathcal {T}$ can then be written as
\begin{equation}
\Theta=dM-TdS-\Phi dQ.\label{21}
\end{equation}%
The non-degenerate metric $G$ from Eq.(\ref{19}) can be changed into
\begin{equation}
G=(dM-TdS-\Phi dQ)^2+(TS+\Phi Q)(-dSdT+dQd\Phi).\label{22}
\end{equation}%
Introducing the map
\begin{equation}
\varphi:\{S,Q\}\mapsto\{M(S,Q),S,Q,\frac{\partial M}{\partial
S},\frac{\partial M}{\partial Q}\},\label{23}
\end{equation}%
the space of thermodynamic equilibrium states can be induced.
According to Eq.(\ref{19}), the thermodynamic metric $g$ can be
written as follows
\begin{equation}
g=(S\frac{\partial M}{\partial S}+Q\frac{\partial M}{\partial
Q})(-\frac{\partial^2M}{\partial S^2}dS^2+\frac{\partial^2
M}{\partial Q^2}dQ^2).\label{24}
\end{equation}%
Utilizing Eqs.(\ref{6}) and (\ref{9}), we can easily calculate the
relevant quantities in Eq.(\ref{24}) as
\begin{align}
\frac{\partial M}{\partial
S}&=\frac{r_+^2+2\tilde{\alpha}-Q^2}{4\pi
r_+(r_+^2-4\tilde{\alpha})},\label{25}
\\
\frac{\partial M}{\partial Q}&=\frac{Q}{r_+},\label{26}
\\
\frac{\partial^2 M}{\partial
S^2}&=\frac{8\tilde{\alpha}^2-r_+^4-10r_+^2\tilde{\alpha}
-Q^2(4\tilde{\alpha}-3r_+^2)}{8\pi^2
r_+(r_+^2-4\tilde{\alpha})^3},\label{27}
\\
\frac{\partial^2 M}{\partial Q^2}&=\frac{1}{r_+}.\label{28}
\end{align}%
Comparing Eqs.(\ref{25}),(\ref{26}) with Eqs.(\ref{7}),(\ref{8}), we
find
\begin{equation}
\frac{\partial M}{\partial S}=T,\quad \frac{\partial M}{\partial
Q}=\Phi,\label{29}
\end{equation}%
which proves the validity of the first law of black hole
thermodynamics $dM=TdS+\Phi dQ$. Substituting
Eqs.(\ref{25})-(\ref{28}) into Eq.(\ref{24}), we can calculate the
component of the thermodynamic metric $g$ as
\begin{align}
g_{SS}=&\frac{1}{32\pi
^2r_+^2(r_+^2-4\tilde{\alpha})^4}\times[r_+^4-8\tilde{\alpha}^2+10r_+^2\tilde{\alpha}+Q^2(4\tilde{\alpha}-3r_+^2)]
\nonumber
\\
&\times[r_+^4+2r_+^2\tilde{\alpha}+Q^2(3r_+^2-16\tilde{\alpha})+8\tilde{\alpha}(Q^2-r_+^2-2\tilde{\alpha})ln{r_+}],\label{30}
\\
g_{QQ}=&\frac{r_+^4-16Q^2\tilde{\alpha}+2r_+^2\tilde{\alpha}+3Q^2r_+^2+(8Q^2\tilde{\alpha}-8r_+^2\tilde{\alpha}-16\tilde{\alpha}^2)ln{r_+}}{4r_+^2(r_+^2-4\tilde{\alpha})}.\label{31}
\end{align}%

Utilizing Eqs.(\ref{30}) and (\ref{31}), we can obtain the Legendre
invariant scalar curvature as
\begin{equation}
\mathfrak{R}_Q=\frac{A(x_+,Q)}{B(x_+,Q)},\label{32}
\end{equation}%
where
\begin{align}
B(x_+,Q)=&[r_+^4+10r_+^2\tilde{\alpha}-8\tilde{\alpha}^2+Q^2(4\tilde{\alpha}-3r_+^2)]^2
\nonumber
\\
&\times[r_+^4+Q^2(3r_+^2-16\tilde{\alpha})+2r_+^2\tilde{\alpha}+8\tilde{\alpha}(Q^2-r_+^2-2\tilde{\alpha})ln{r_+}]^3
 .\label{33}
\end{align}%

\begin{figure}
\includegraphics[width=8cm,height=6cm]{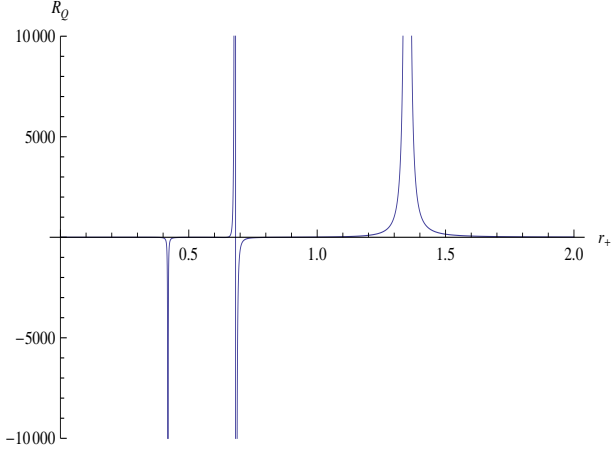}
 \caption{Thermodynamic scalar curvature $R _Q$ vs. $r_+$ for $Q=1,\tilde{\alpha}=0.1$}
\label{fg5}
\end{figure}

The numerator of the Legendre invariant scalar curvature is too
lengthy to be displayed here. From Eq.(\ref{33}), we find that the
Legendre invariant scalar curvature shares the same factor
$r_+^4+10r_+^2\tilde{\alpha}-8\tilde{\alpha}^2+Q^2(4\tilde{\alpha}-3r_+^2)$
with the specific heat $C_Q$ in its denominator , which implies that
it would diverge when
$r_+^4+10r_+^2\tilde{\alpha}-8\tilde{\alpha}^2+Q^2(4\tilde{\alpha}-3r_+^2)=0$.
That is the exact condition that the phase transition point
satisfies. To get an intuitive sense on this issue, we plot
Fig.\ref{fg5} showing the behavior of thermodynamic scalar curvature
$\mathfrak{R}_Q$. From Fig.\ref{fg5}, we find that thermodynamic
scalar curvature $\mathfrak{R}_Q$ diverges at three locations.
Comparing Fig.\ref{fg5} with Fig.\ref{1b}, we find that the second
diverging point which corresponds to negative Hawking temperature
does not have physical meaning. Furthermore, the first and the third
diverging points coincide exactly with the phase transition point,
which can be witnessed by comparing Fig.\ref{fg5} with Fig.\ref{1a}.
So we can safely draw the conclusion that the Legendre invariant
metric constructed in geometrothermodynamics correctly produces the
behavior of the thermodynamic interaction and phase transition
structure of black holes with conformal anomaly.

\section{Critical exponents and scaling laws}
 \label{sec:5}
 In order to have a better understanding of the phase transition of
 black holes with conformal anomaly, we would like to investigate
 their critical behavior near the critical point by considering a
 set of critical exponents in this section.

 Before embarking on calculating critical exponents, we would like to reexpress physical quantities near the critical point as
\begin{align}
r_+&=r_c(1+\Delta),\label{34}
\\
T(r_+)&=T_c(1+\varepsilon) ,\label{35}
\\
Q(r_+)&=Q_c(1+\eta) ,\label{36}
\end{align}%
where $|\Delta|\ll 1, |\varepsilon|\ll 1, |\eta|\ll 1$. Note that
the footnote "c" in this section
 denotes the value of the physical quantity (or the expression) at the critical point.
 For example, $T_c$ corresponds to the temperature at the critical
 point.

 Critical
exponent $\alpha$ is defined through
\begin{equation}
C_Q\sim|T-T_c|^{-\alpha}.\label{37}
\end{equation}%
To obtain $T-T_c$, we would like to carry out Taylor expansion as
below
\begin{equation}
T(r_+)=T_c+[(\frac{\partial T}{\partial
r_+})_{Q=Q_c}]_{r_+=r_c}(r_+-r_c)+\frac{1}{2}[(\frac{\partial ^2
T}{\partial r_+ ^2})_{Q=Q_c}]_{r_+=r_c}(r_+-r_c)^2+higher\, order\,
terms, \label{38}
\end{equation}%
from which we obtain
\begin{equation}
\Delta=\frac{1}{r_c}\sqrt{\frac{2 \varepsilon T_c}{D}},\label{39}
\end{equation}%
where
\begin{equation}
D=[(\frac{\partial ^2 T}{\partial r_+
^2})_{Q=Q_c}]_{r_+=r_c}=\frac{r_c^6+24r_c^4\tilde{\alpha}-24r_c^2\tilde{\alpha}^2+32\tilde{\alpha}^3-2Q_c^2(3r_c^4-6r_c^2\tilde{\alpha}+8\tilde{\alpha}^2)}{2\pi
(r_c^3-4r_c\tilde{\alpha})^3}.\label{40}
\end{equation}%
In the above derivation, we have considered the fact that $C_Q$
diverges at the critical point, making the second term at the right
hand side of Eq.(\ref{38}) vanish. Substituting Eq.(\ref{34}) into
Eq.(\ref{10}) and keeping only the linear terms in its denominator,
we obtain
\begin{equation}
C_Q\simeq\frac{2\pi(r_c^2-4\tilde{\alpha})^2(Q_c^2-2\tilde{\alpha}-r_c^2)}{\Delta(4r_c^4+20r_c^2\tilde{\alpha}-6Q_c^2r_c^2)},\label{41}
\end{equation}%
which can be transformed via Eq.(\ref{39}) into
\begin{equation}
C_Q\simeq\frac{\pi \sqrt{2D}
(r_c^2-4\tilde{\alpha})^2(Q_c^2-2\tilde{\alpha}-r_c^2)}{(4r_c^3+20r_c\tilde{\alpha}-6Q_c^2r_c)(T-T_c)^{1/2}},\label{42}
\end{equation}%
Comparing Eq.(\ref{42}) with Eq.(\ref{37}), we can obtain
$\alpha=1/2$.

 Critical
exponent $\beta$ is defined through the following relation when $Q$
is fixed,
\begin{equation}
\Phi(r_+)-\Phi(r_c)\sim|T-T_c|^{\beta}.\label{43}
\end{equation}%
The above definition motivates us to carry out the Taylor expansion
as
\begin{equation}
\Phi(r_+)=\Phi_c+[(\frac{\partial \Phi}{\partial
r_+})_{Q=Q_c}]_{r_+=r_c}(r_+-r_c)+higher\, order\, terms,\label{44}
\end{equation}%
Utilizing Eq.(\ref{8}), (\ref{44}) and neglecting higher order terms
of Eq.(\ref{43}), we get
\begin{equation}
\Phi(r_+)-\Phi_c=[(\frac{\partial \Phi}{\partial
r_+})_{Q=Q_c}]_{r_+=r_c}\sqrt{\frac{2}{D}}(T-T_c)^{1/2}=-\frac{Q_c}{r_c^2}\sqrt{\frac{2}{D}}(T-T_c)^{1/2}.\label{45}
\end{equation}%
Comparing Eq.(\ref{43}) with Eq.(\ref{45}), we can obtain
$\beta=1/2$.

 Critical
exponent $\gamma$ is defined through the following relation
\begin{equation}
\kappa_T^{-1}\sim|T-T_c|^{-\gamma}.\label{46}
\end{equation}%
Substituting Eq.(\ref{34}) and (\ref{39}) into Eq.(\ref{18}) and
keeping only the linear term of $\Delta$, we obtain
\begin{equation}
\kappa
_T^{-1}=\frac{\sqrt{D}(Q_cr_c^4-Q_c^3r_c^2-4Q_c^3\tilde{\alpha}+10Q_cr_c^2\tilde{\alpha}-8Q_c\tilde{\alpha}^2)}{\sqrt{2}[5r_c^4-8\tilde{\alpha}^2+30r_c^2\tilde{\alpha}
+Q_c^2(4\tilde{\alpha}-9r_c^2)](T-T_c)^{\frac{1}{2}}}.\label{47}
\end{equation}
From Eq.(\ref{46}) and (\ref{47}), we find that $\gamma=1/2$

 Critical
exponent $\delta$ is defined for the fixed temperature $T_c$ through
\begin{equation}
\Phi(r_+)-\Phi(r_c)\sim|Q-Q_c|^{1/\delta}.\label{48}
\end{equation}%
To obtain $Q-Q_c$, we would like to carry out Taylor expansion as
\begin{equation}
Q(r_+)=Q_c+[(\frac{\partial Q}{\partial
r_+})_T]_{r_+=r_c}(r_+-r_c)+\frac{1}{2}[(\frac{\partial ^2
Q}{\partial r_+ ^2})_T]_{r_+=r_c}(r_+-r_c)^2+higher\, order\, terms,
\label{49}
\end{equation}%
Utilizing the thermodynamic identity again, we get
\begin{equation}
[(\frac{\partial Q}{\partial r_+})_T]_{r_+=r_c}=-[(\frac{\partial
T}{\partial r_+})_Q]_{r_+=r_c}[(\frac{\partial Q}{\partial
T})_{r_+}]_{r_+=r_c}=0.\label{50}
\end{equation}
In the above derivation, we have taken into account the fact that
$C_Q$ diverges at the critical point, making the first term at the
right hand side of Eq.(\ref{38}) vanish. Substituting Eq.(\ref{34})
and Eq.(\ref{36}) into Eq.(\ref{49})and neglecting the high order
terms, we obtain
\begin{equation}
\Delta=\sqrt{\frac{2Q_c \eta}{E r_c^2}},\label{51}
\end{equation}
where
\begin{equation}
E=[(\frac{\partial ^2 Q}{\partial
r_+^2})_T]_{r_+=r_c}=\frac{22r_c^4\tilde{\alpha}+32\tilde{\alpha}^3+16\tilde{\alpha}^2(r_c^2-Q_c^2)-r_c^4(3Q_c^2+r_c^2)}{2Q_c(r_c^3-4r_c\tilde{\alpha})^2}.\label{52}
\end{equation}
Taylor expanding $\Phi$ near the critical point, we get
\begin{equation}
\Phi(r_+)=\Phi_c+[(\frac{\partial \Phi}{\partial
r_+})_T]_{r_+=r_c}(r_+-r_c)+higher\, order\, terms, \label{53}
\end{equation}%
where the coefficient of the second term on the right hand side can
be derived as follows
\begin{equation}
[(\frac{\partial \Phi}{\partial r_+})_T]_{r_+=r_c}=[(\frac{\partial
Q}{\partial r_+})_T]_{r_+=r_c}[(\frac{\partial \Phi}{\partial
Q})_{r_+}]_{r_+=r_c}+[(\frac{\partial \Phi}{\partial
r_+})_Q]_{r_+=r_c}=-\frac{Q_c}{r_c^2}.\label{54}
\end{equation}
Utilizing Eq.(\ref{51}), (\ref{53}), (\ref{54}),we get
\begin{equation}
\Phi(r_+)-\Phi_c\simeq-\frac{Q_c}{r_c^2} \sqrt{\frac{2(Q-Q_c)}{E}},
\label{55}
\end{equation}%
from which we can draw the conclusion that $\delta=2$.

Critical exponent $\varphi$ is defined through
\begin{equation}
C_Q\sim|Q-Q_c|^{-\varphi}.\label{56}
\end{equation}%
Substituting Eq.(\ref{51}) into Eq.(\ref{41}), we obtain
\begin{equation}
C_Q\simeq\frac{\pi r_c
\sqrt{2E}(r_c^2-4\tilde{\alpha})^2(Q_c^2-2\tilde{\alpha}-r_c^2)}{\sqrt{Q-Q_c}(4r_c^4+20r_c^2\tilde{\alpha}-6Q_c^2r_c^2)},\label{57}
\end{equation}%
Comparing Eq.(\ref{57}) and (\ref{56}), we find that $\varphi=1/2$.

Critical exponent $\psi$ is defined through
\begin{equation}
S(r_+)-S_c\sim|Q-Q_c|^{\psi}.\label{58}
\end{equation}%
Taylor expanding $S$ near the critical point, we obtain
\begin{equation}
S(r_+)=S_c+[(\frac{\partial S}{\partial
r_+})_Q]_{r_+=r_c}(r_+-r_c)+higher\, order\, terms. \label{59}
\end{equation}%
Utilizing Eq.(\ref{9}), (\ref{34}), (\ref{51}) and (\ref{59}), we
get
\begin{equation}
S(r_+)-S_c\simeq(2\pi r_c-\frac{8\pi
\tilde{\alpha}}{r_c})\sqrt{\frac{2(Q-Q_c)}{E}}, \label{60}
\end{equation}%
from which we obtain $\psi=1/2$.

Till now, we have finished the calculations of six critical
exponents. They are also equal to $1/2$ except $\delta=2$. Our
results are in accordance with those in classical thermodynamics.
And it can be easily proved that critical exponents we obtain in our
paper satisfy the following thermodynamic scaling laws
\begin{eqnarray}
\alpha+2\beta+\gamma=2,\, \alpha+\beta(\delta+1)=2,\,
(2-\alpha)(\delta \psi-1)+1=(1-\alpha)\delta, \nonumber
\\
\gamma(\delta+1)=(2-\alpha)(\delta-1),\,\gamma=\beta(\delta-1),\,\varphi+2\psi-\delta^{-1}=1.\label{61}
\end{eqnarray}%

\section{Conclusions}
\label{sec:6}
  The phase transition of black holes with conformal anomaly has been investigated in canonical ensemble. Firstly, we calculate the relevant thermodynamic quantities and discuss the
  behavior of the specific heat at constant charge. We find that there have been striking
  differences
between black holes with conformal anomaly and those without
conformal anomaly. In the case $Q=1,\tilde{\alpha}=0.1$, there are
two phase transition point while there is only one in the case
$\tilde{\alpha}=0$. The temperature in the case $\tilde{\alpha}=0$
increases monotonically while there exists local minimum temperature
in the case $Q=1,\tilde{\alpha}=0.1$. This local minimum temperature
corresponds to the phase transition point. We also find that the
phase transitions of black holes with conformal anomaly take place
not only from an unstable large black hole to a locally stable
medium black hole but also from an unstable medium black hole to a
locally stable small black hole. We also study the behavior of the
inverse of the isothermal compressibility $\kappa_T^{-1}$ and find
that $\kappa_T^{-1}$ also diverges at the critical point.

Secondly, we probe the dependence of phase transitions on the choice
of parameters. The results show that black holes with conformal
anomaly have much richer phase structure than that without conformal
anomaly. When $\tilde{\alpha}=0$, the location of the phase
transition $r_c$ is proportional to the charge $Q$. By contrast, the
case of black holes with conformal anomaly is more complicated. For
$\tilde{\alpha}=0.1$, the curve can be divided into three regions.
Through numerical calculation, we find that black holes has only one
phase transition point when $Q\subset(0,0.4472)$. When
$0.4472<Q<0.7746$, there would be no phase transition at all. When
$Q>0.7746$, there exist two phase transition points. And the
distance between these two phase transition points becomes larger
with the increasing of $Q$. In the case that the charge $Q$ has been
fixed at one, we notice that there would be two phase transition
point when $0<\tilde{\alpha}<\frac{1}{6}$. And the distance between
these two phase transition points becomes narrower with the
increasing of $\tilde{\alpha}$. When
$\tilde{\alpha}\subset(\frac{1}{6},\frac{1}{2})$, there would be no
phase transition. When $\tilde{\alpha}>\frac{1}{2}$, there would be
only one phase transition point.

  Thirdly, we build up geometrothermodynamics in canonical ensembles. We
  choose $M$ to be the thermodynamic potential and build up both thermodynamic phase space and the space of thermodynamic equilibrium
  states. We also calculate the Legendre invariant thermodynamic scalar
  curvature and depict its behavior graphically. It is shown that Legendre invariant thermodynamic scalar
  curvature diverges exactly where the specific heat diverges. Based on this, we can safely conclude that the Legendre
invariant metrics constructed in geometrothermodynamics can
correctly produce the behavior of the thermodynamic interaction and
phase transition structure even when conformal anomaly is taken into
account.

Furthermore, we calculate the relevant critical exponents. They are
also equal to $1/2$ except $\delta=2$. Our results are in accordance
with those of other black holes. And it has been proved that
critical exponents we obtain in our paper satisfy the
thermodynamic scaling laws. We conclude that the critical exponents
and the scaling laws do not change even when we consider conformal
anomaly. This may be attributed to the mean field theory.

\acknowledgments This research is supported by the National Natural Science
Foundation of China (Grant Nos.11235003, 11175019, 11178007). It is
also supported by "Thousand Hundred Ten" project of Guangdong
Province and Natural Science Foundation of Zhanjiang Normal
University under Grant No. QL1104.

\end{document}